\documentstyle[aps,preprint]{revtex}

\begin{document}

\title{
CHIRAL MESON MASSES AT FINITE TEMPERATURE AND DENSITY}
\author{
A. Barducci, R. Casalbuoni, G. Pettini}
\address{Dipartimento di Fisica, Universit\`a di Firenze,
I-50125 Firenze, Italy\\
and Istituto Nazionale di Fisica Nucleare, Sezione di Firenze,
I-50125 Firenze, Italy}
\author{R. Gatto}
\address{D\'epartement de Physique Th\'eorique, Universit\'e de Gen\`eve,
CH-1211 Gen\`eve 4, Switzerland}

\maketitle
\vspace{0.5cm}
\centerline{UGVA-DPT 2000/08-1087}
\centerline{Firenze Preprint - DFF - 359/06/2000}
\vspace{-0.5cm}

\draft

\begin{abstract}

The ratio of the sigma mass to the pion mass at finite temperatures and
densities
provides for a quantitative signal of  chiral symmetry breaking. We calculate
this ratio
by using an extension to finite chemical potential of the field theoretic
composite operator formalism as applied to QCD. The calculation is limited
to regions of the phase diagram where only quark-antiquark condensates
dominate  (no quark-quark condensates) and it confirms the expected
behaviours. In particular the sigma becomes an essentially stable particle
in a narrow region bordering the transition line from broken to restored
chirality . This pattern is qualitatively the same both for the region
where the transition is of second order as for the region where it is of
first order
(apart from the discontinuities expected in the latter case).

\end{abstract}



\section{Introduction}

Since the original discussion by Nambu \cite{nambu} in 1960, chiral
symmetry breaking has played
a crucial role in the theory  of strong interactions. In 1974  Lee and Wick
\cite{wick}
proposed that at high densities chiral symmetry may be restored.
Restoration at high temperatures
was discussed in the same period by Dolan and Jackiw \cite{dolan} and by
Weinberg \cite{weinberg}.
Condensation of quark-antiquark pairs plays the main role in chiral
symmetry breaking, except for
very high densities where quark-quark condensates are expected to play the
important role.
In this note we do not consider this very high density region and limit
ourselves to densities
such that only quark-antiquark condensates are important. Such a region is
the one important in
hadronic physics, and chiral restoration from this region is expected to be
reachable with high energy heavy ion collisions \cite{satz}. Chiral
transition through the
quark-antiquark  order parameter is also expected to be of great  relevance
for the physics
of the early universe \cite{linde}.

A great number of studies has been carried out on the
expected QCD phase diagram (see for instance \cite{phase}).  For a
review of the subject see the recent paper by Rajagopal \cite{Rajago} 

 Lattice
calculations
are rapidly progressing and they will with time provide for a complete
approach
(a review  can be found for instance in Ref.
\cite{karsch}). Many theoretical approaches use the Nambu-Jona-Lasinio model
and  different dynamical approximations for quantitative indications on the
chiral transition
(see for instance \cite{bernard}).

Besides the chiral transition also the deconfinement transition takes
place. A single
transition is generally expected to occur, rather than separate transitions
for deconfinement and
chiral-symmetry (see for instance for a heuristic argument Ref
\cite{heuristic}). Theoretically
one has to deal with different order parameters. Moreover they cover
extreme ranges in opposite
domains.
The transition from the confined to the non-confined phase is often
described in terms of
the thermally averaged Polyakov loop which applies particularly in the
limit of
infinite quark masses. For the chiral symmetry transition the typical order
parameters are the thermally averaged quark-antiquark bilinears (at least
up to not
very high densities) which apply mostly in the limit of vanishing quark
masses.

The current quark mass violates explicitly  the chiral symmetry whose
restoration
characterizes the phase transition. Its  role is formally analogous to that
of the
external magnetic field in the ferromagnetic transition. As long as the
mass is small, such as
for $u$ and $d$ quarks, the notion of phase transition can be approximately
useful.

Our study
uses a composite operator formalism for both finite temperature and
finite density. It is
based  on constructing the effective action for the composite
operators \cite{bcd,masst} in the presence of temperature and chemical
potential. We focus
on the study of the ratio of the scalar to pseudoscalar mass
$M_{\sigma}/M_{\pi}$ in the plane of
temperature and chemical potential. The ratio might have a physical
interest as a possible
experimental signature for the transition, although it is too early for
assessing its usefulness in
these regards. It has anyway a theoretical interest as it constitutes a
significant parameter for the degree of chiral symmetry breaking at
finite temperatures and densities. The ratio is equal to unity when chiral
symmetry is restored
(we neglect for the moment the small $u$ and $d$ current quark masses).
In the hadronic phase,
at zero temperature and small density, in the broken chirality phase, the
pion is a goldstone
of vanishing mass, while the sigma has a finite mass from the chiral
condensation.
The ratio of the scalar to the pseudoscalar mass  is thus
expected  to decrease from very large values to a value of one at the chiral
transition (the $u$ and $d$ current quark masses only slightly change
this picture).

A physical consequence of this behaviour is that, before approaching
the critical values
for the transition, by increasing together or separately temperature and
density,
for the sigma particle there is not any phase space left for its decay
into two pions.
Its main source of instability is thus suppressed  when approaching
the phase border in
the two dimensional plane of temperature and chemical potential.
Consequently the sigma
becomes an almost stable particle with dominant decay channel
\cite{pisarski} that
into two photons. The suggested experimental signature would thus be in that
region a bump
in the invariant mass of two photons. We do not know at this stage how this
will be visible in
heavy ion collisions.

On the other hand it is worthwhile to recall that indications on the sigma
at zero temperature
and density have been rather elusive just because of the very large width
the sigma has in that
case. It has not been demonstrated so far that the pion-pion phase shift
passes through 90 degrees in a
region below some 700-800 MeV, suggesting that the large width may mask the
overall effect. Also it
is well-known that strong rescattering effects in different channels make
the interpretation of the
low energy phenomenology in the $0^{++}$ channel not easy \cite{rescattering}.
We just mention that a recent experiment by the E91 collaboration brings
evidence for a light and broad
scalar resonance in the D-decay channel into three pions \cite{aitala} and
that a simple model makes
plausible the interpretation in terms of a sigma resonance \cite{GNPT}. For
further phenomenological
discussions on the sigma resonance in hadronic physics we refer to the papers
posted in relation to
the recent conference on "Possible existence of a light sigma resonance and
its implications to
hadron physics", held in Kyoto, June 11-14 (for a review see the summary talk
by Tornqvist \cite{Tornqvist}). We also mention that there is been
recently discussion as to the possibility of a glueball interpretation 
of the sigma resonance and for this we refer to a recent talk by
Pennington disfavoring such an interpretation \cite{Pennington}.  

We recall that the QCD phase diagram, even for massless quarks, is known to
have a richer
structure when studied in the plane of T and chemical potential, as one
expects both types
of transitions, of second order for low chemical potentials and of first
order beyond some critical
value of the chemical potential. Besides, for large chemical potential a
color superconductivity
phase should occur, but here we limit ourselves to consequences of
quark-antiquark condensations
and assume not yet to be in a region where quark-quark condensates appear.
The separation along
the critical line between the two types of transitions defines a
tricritical point which has
been the subject of extensive studies \cite{tricritical,tricritical1}. As
far as the
particular subproduct
of the present calculation, that is finding the region in the phase diagram
where the sigma becomes
of such a low mass to be no longer able to decay into two pions and thus
appearing as a
narrow state, we find that, independently on the type of transition,
this region is anyway
confined to a small strip near the transition line. What changes of course
is that when passing
through the transition line one finds for the the ratio of the sigma to the
pion mass a discontinuity
in the case when the transition is of first order.

In Section II we recall the main steps to evaluate the effective potential
in the composite operator formalism. Then we derive the expressions for
the pion and $\sigma$-meson masses at finite temperature and density in our
model. Finally we obtain a formula for their ratio and we show the numerical
results for typical behaviours when crossing the lines of second
and first order phase transitions.

\section{Effective action}
\label{sec:effact}

 The zero temperature Euclidean
effective action for an $SU(N)$ QCD-like gauge theory is (we follow
\cite{bcd})
\begin{equation}
\Gamma( {\bf\Sigma} ) = - {\rm Tr}\ln\left[{\bf S}_0^{-1} + {\delta
\Gamma_2 \over\delta{\bf S}}\right] - {\rm Tr}\left[
{\bf \Sigma}{\bf S}\right] -\Gamma_2({\bf S})
+ counterterms
\label{effact}
\end{equation}
where
\begin{equation}
{\bf S}^{-1}_0= (i\hat{p}- {\bf m})
\label{two}
\end{equation}
{\bf m} is the bare quark mass matrix and $\Gamma_2({\bf S})$
is the sum of all the
two-particle irreducible vacuum diagrams with fermionic propagator
${\bf S}$ and ${\bf \Sigma} = -\delta \Gamma_2/\delta{\bf S}$.
One has at two-loop level
\begin{equation}
\Gamma_2={1\over2}{\rm Tr}({\bf S}\Delta{\bf S})
\label{three}
\end{equation}
where $\Delta$ is the gauge boson propagator, so that
\begin{equation}
{\bf \Sigma} = - \Delta{\bf S},
\label{four}
\end{equation}

One can therefore rewrite Eq.\ (\ref{effact})
\begin{eqnarray}
\Gamma( {\bf\Sigma} ) &=& - {\rm Tr}\ln\left[{\bf S}_0^{-1} -{\bf \Sigma}
\right] + {1\over 2}{\rm Tr}\left({\bf \Sigma}\Delta^{-1}{\bf \Sigma}
\right) + counterterms
\label{effact1}
\end{eqnarray}

A parametrization for ${\bf \Sigma}$ employed in \cite{bcd} was
\begin{equation}
 {\bf\Sigma}=({\bf s} + i \gamma_5{\bf p})f(k)\equiv {\bf\Sigma}_s
+i\gamma_5 {\bf\Sigma}_p
\label{selfen}
\end{equation}
 with {\bf s} and {\bf p}
scalar and pseudoscalar constant fields respectively and with a suitable
Ansatz for $f(k)$.

  From Eq.\ (\ref{effact1})
(see Ref. \cite{bcd}) one obtains the effective potential
\begin{eqnarray}
V={\Gamma\over\Omega}&=& -{8\pi^2 N\over 3 C_2 g^2}
\int{d^4 k\over(2\pi)^4}{\rm tr}\left[{\bf \Sigma}_s\Box_k{\bf \Sigma}_s
+{\bf \Sigma}_p\Box_k{\bf \Sigma}_p\right]-\nonumber\\
&&- N~{\rm Tr}\ln\left[i\hat{k} - \left({\bf m}+{\bf \Sigma}_s
\right)-i\gamma_5{\bf \Sigma}_p\right] + \delta Z~{\rm tr}({\bf m}~{\bf s})
\label{effpot}
\end{eqnarray}
where $C_2$ is the quadratic Casimir of the fermion representation
(for $SU(3)_{c}$  $C_{2}=4/3$), $g$ is the gauge
coupling constant and $\Omega$ the four-dimensional volume.
Furthermore
\begin{eqnarray}
{\bf \Sigma}_s &\equiv&
{\lambda_{\alpha}\over \sqrt{2}}~ s_{\alpha}~ f(k)\cr
{\bf \Sigma}_p &\equiv&{\lambda_{\alpha}\over \sqrt{2}}~ p_{\alpha}~ f(k)\cr
{\bf m}&\equiv&{\lambda_{\alpha}\over\sqrt{2}}~m_\alpha
\end{eqnarray}
\label{eight}

with $\alpha=0,\cdots,8$;~~$\lambda_0=\sqrt{2/3}~I$~~ and $\lambda_i$~
($i=1,\cdots,8$) are the Gell-Mann matrices.

For $f(k)$ we have chosen as a variational Ansatz \cite{tricritical}
\begin{equation}
f(k)={M^{3}\over
 k^2+M^2}
\label{nine}
\end{equation}
where $M$ is a momentum scale expected to be of the order
of $\Lambda_{QCD}$.

The extension to non zero temperature $T$ and non zero chemical
potential $\mu$ corrects the effective potential for a quark of mass $m$ to
the form
\begin{eqnarray}
V={\Gamma\over\Omega}&=& -{8\pi^2 N\over 3 C_2 g^2(T,\mu)}
\int{d^4 k\over(2\pi)^4}\left[\Sigma_s\Box_k\Sigma_s
+\Sigma_p\Box_k\Sigma_p\right]-\nonumber\\
&&- 2N~\sum_{n=-\infty}^{+\infty}
(-)^{n}\int{d^4 k\over(2\pi)^4}
\ln\left[{\bar k}^2+\left(m+\Sigma_s\right)^2
+\Sigma_{p}^{2}\right]~e^{\displaystyle{i~n{k_{0}\over T}}}
~+~ \delta Z~ m~ s
\label{eleven}
\end{eqnarray}
where ${\bar k}^{\nu}\equiv(k_{0}+i\mu,{\bf k})$
and we have used the Poisson formula, equivalent
to the application of standard imaginary-times rules
\cite{dolan},\cite{tricritical,solot}.

In Eq.\ (\ref{eleven})
\begin{equation}
\Sigma_{s}~=~s~f({\bar k})
\label{twelve}
\end{equation}
and analogously for $\Sigma_{p}$.
The coupling constant depends on $T$ and $\mu$ as
\begin{equation}
{2\pi^{2}\over g^{2}(T,\mu)}\equiv c(T,\mu)\equiv
c_{0}+c_{1}(T,\mu)=c_{0}~+~{\pi^{2}\over b}~
{\rm ln}\left(1+\xi {T^{2}\over M^{2}}+\zeta {\mu^{2}\over M^{2}}
\right)
\label{thirteen}
\end{equation}
with $b=24\pi^2/(11N-2n_{f})$, and we will discuss the determination
of the parameters $c_{0},M,\xi$ and $\zeta$ later on.
Also for non vanishing $T$ and $\mu$ we impose the normalization
\begin{equation}
\lim_{m \rightarrow 0}~ {\partial V\over \partial
\left( m\langle {\bar \psi}\psi\rangle_{0}\right)}
\Big|_{{\rm min}}=1
\label{fourteen}
\end{equation}

The following relation relates the minimum of the potential ${\bar s}(T,\mu)$
to the condensates $\langle {\bar \psi}\psi\rangle_{T,\mu}$

\begin{equation}
\langle {\bar \psi}\psi\rangle_{T,\mu}={N~M^{3}\over g^{2}(T,\mu)}
~{\bar s}(T,\mu)
\label{fifteen}
\end{equation}

When using physical pion fields with their proper normalization, the
normalization condition of
Eq.\ (\ref{fourteen}) is equivalent to the Adler-Dashen formula

\begin{equation}
M^{2}_{\pi}(T,\mu)f^{2}_{\pi}(T,\mu)~=
~-2m\langle {\bar \psi}\psi\rangle_{T,\mu}
\label{sixteen}
\end{equation}

\section{Masses}
\label{sec:masse}

To derive the masses we use canonical fields

\begin{eqnarray}
\varphi_{\sigma}~&=&~a~s\cr
\varphi_{\pi}~&=&~a~p\cr
a~&=&~-{f_{\pi}\over\sqrt{2} {\bar s}}
\label{seventeen}
\end{eqnarray}

and we obtain

\begin{equation}
M^{2}_{\sigma}={\partial^2 V\over \partial\varphi_{\sigma}^{2}}
\Big|_{\rm min}={1\over a^2}{\partial^2 V\over \partial s^{2}}
\Big|_{\rm min}=2{{\bar s}^2\over f_{\pi}^2}{\partial^2 V\over
\partial s^{2}}\Big|_{{\rm min}}
\label{eighteena}
\end{equation}

\begin{equation}
M^{2}_{\pi}={\partial^2 V\over \partial\varphi_{\pi}^{2}}
\Big|_{\rm min}={1\over a^2}{\partial^2 V\over \partial p^{2}}
\Big|_{\rm min}=2{{\bar s}^2\over f_{\pi}^2}{\partial^2 V\over
\partial p^{2}}\Big|_{{\rm min}}
\label{eighteenb}
\end{equation}

where ${\bar s}$ is the minimum of the effective potential
in presence of the quark bare mass $m$.

In the following we use the minimum condition
\begin{eqnarray}
{\partial V\over\partial s}=0\rightarrow &&\cr
{N\over {\bar s}}\Big[&-&2c(T,\mu)\int{d^4 k\over (2\pi)^4}
{\bar\Sigma_s}\Box_k{\bar \Sigma_s}\cr
&-&4\sum_{n=-\infty}^{+\infty}(-)^{n}\int{d^4 k\over (2\pi)^4}
{\left( m+{\bar \Sigma}_s\right){\bar \Sigma}_s\over
\left[{\bar k}^2+\left( m+{\bar \Sigma}_s\right)^2+{\bar \Sigma}_{p}^{2}
\right]}e^{\displaystyle{in{k_{0}\over T}}}+m\delta Z\Big]=0\cr
{\partial V\over\partial p}=0\rightarrow && {\bar p}=0
\label{nineteen}
\end{eqnarray}

where ${\bar\Sigma_s}\equiv {\bar s}f({\bar k})$.
Furthermore we have the normalization condition for
the effective potential

\begin{equation}
\delta Z=N\left[{M^3\over 2\pi^2}c(T,\mu)+
{4\over s_{0}}\sum_{n=-\infty}^{+\infty}(-)^{n}
\int{d^4 k\over (2\pi)^4}
{\Sigma_{0}\over {\bar k}^2+\Sigma_{0}^2}
e^{\displaystyle{in{k_{0}\over T}}}\right]
\label{twenty}
\end{equation}

with  $\Sigma_{0}\equiv s_{0}f({\bar k})$ and $s_{0}$ is the minimum
for the case of a massless quark. The two preceding equations
allow for the following expressions for the masses

\begin{eqnarray}
M_{\pi}^{2}(T,\mu)=&-&2{m\over f_{\pi}^{2}(T,\mu)}
\langle{\bar\psi}\psi\rangle_{T,\mu}+\cr
&+&8N{m\over f_{\pi}^{2}(T,\mu)}\sum_{n=-\infty}^{+\infty}(-)^{n}
\int{d^4 k\over (2\pi)^4}\Big[{{\bar\Sigma}_{s}\over {\bar k}^2+
\left(m+{\bar\Sigma}_{s}\right)^2}-\cr
&-&{{\bar s}(T,\mu)\over s_{0}(T,\mu)}
{\Sigma_{0}\over {\bar k}^2+\Sigma_{0}^2}\Big]
e^{\displaystyle{in{k_{0}\over T}}}
\label{twentyonea}
\end{eqnarray}

\begin{equation}
M_{\sigma}^{2}(T,\mu)=M_{\pi}^{2}(T,\mu)+{16N\over f_{\pi}^{2}(T,\mu)}
\sum_{n=-\infty}^{+\infty}(-)^{n}
\int{d^4 k\over (2\pi)^4}{\left(m+{\bar\Sigma}_{s}\right)^2
\over \left[{\bar k}^2+
\left(m+{\bar\Sigma}_{s}\right)^2\right]^{2}}~{\bar\Sigma}_{s}^2
~e^{\displaystyle{in{k_{0}\over T}}}
\label{twentyoneb}
\end{equation}

The parameters $c_{0}$ and $s_{0}(T=\mu=0)$ are fixed by following
the procedure of Ref. \cite{bcd,masst,tricritical,tokyo} where one was able
to fix the value $c_{0}=0.554$ from the renormalization
condition and the gap equation for the effective potential
in the limit of small current quark masses.
By inserting this value for $c_{0}$ in the gap equation for
massless quarks, one finds that chiral symmetry is spontaneously
broken (at $T=\mu=0$). Actually the origin of $V$ is a local
maximum whereas the absolute minimum is at $s_{0}(T=\mu=0)=-4.06$.
To determine the mass scale $M$ and the quark masses from
the experimental data, one has to derive the explicit expressions for the
masses and for the decay constants of the pseudoscalar octet mesons which
are the pseudo-Goldstone bosons of chiral symmetry breaking.
These expressions constitute a system of coupled equations which
we have solved by an approximation method.
The experimental inputs are the decay constant and mass of the
charged pion, the charged kaon mass and the electromagnetic
mass difference between the neutral and the charged kaon. The
outputs of the numerical fit for the octet meson masses (agreement
within 3\%) are the masses of the $u$,$d$ and $s$ quarks and the mass
parameter $M$. The values we get are the following
\cite{bcd,masst,tokyo}
\begin{eqnarray}
M~&=&~280~MeV\nonumber\\
m_{u}~&=&~8~MeV\nonumber\\
m_{d}~&=&~11~MeV\label{qmasval}\\
m_{s}~&=&~181~MeV\nonumber\\
m~&\equiv&~{m_{u}+m_{d}\over 2}=9.5~MeV\nonumber
\end{eqnarray}
Finally, the parameter $\xi$ in the expression
(\ref{thirteen}) for the running coupling constant has been
set to $\xi\simeq 1$ \cite{fpit} by comparing our model in the
low-$T$ and $\mu=0$
regime with the results of Ref. \cite{leut}, whereas $\zeta$
is left as undetermined \cite{tricritical,masstmu}.

With these values for the relevant parameters it is easy to
determine the masses of the pseudoscalar and scalar mesons at $T=\mu=0$
(we are using the value $f_{\pi}=93~MeV$ for the pion decay constant)
\begin{eqnarray}
M_{\pi}&=&135~MeV\nonumber\\
M_{\sigma}&=&630~MeV
\label{masspisi}
\end{eqnarray}
and (again at $T=\mu=0$) the fermionic condensates at $M$ for massless
and massive quarks
\begin{eqnarray}
\langle{\bar\psi}\psi\rangle&\simeq&(-197~MeV)^3\nonumber\\
\langle{\bar\psi}\psi\rangle&\simeq&(-200~MeV)^3
\label{conds}
\end{eqnarray}
respectively.

\section{The ratio $M_\sigma^2/M_\pi^2$}
\label{sec:rapporto}

From Eq.\ (\ref{twentyoneb}) we can calculate the ratio
$M_{\sigma}^{2}/M_{\pi}^{2}$. We then use the expression for
$f_{\pi}^{2}M_{\pi}^{2}$ obtained from Eq.\ (\ref{twentyonea}), and get

\begin{equation}
{M_{\sigma}^{2}\over M_{\pi}^{2}}=1-{8N\over m}
{
\displaystyle{
\sum_{n=-\infty}^{+\infty}(-)^{n}\int{d^4 k\over (2\pi)^4}
{
\left(m+{\bar\Sigma}_{s}\right)^2~{\bar\Sigma}_{s}^{2}
\over
\left[
{\bar k}^{2}+\left(m+{\bar\Sigma}_{s}\right)^2
\right]^{2}
}
}
~e^{\displaystyle{in{k_{0}\over T}}}
\over
\langle{\bar\psi}\psi\rangle_{T,\mu}
-4N\displaystyle{
\sum_{n=-\infty}^{+\infty}(-)^{n}\int{d^4 k\over (2\pi)^4}
\left[{{\bar\Sigma}_{s}\over
{\bar k}^{2}+\left(m+{\bar\Sigma}_{s}\right)^2}
-{{\bar s}\over s_{0}}{\Sigma_{0}\over {\bar k}^{2}+\Sigma_{0}^2}\right]
}
~e^{\displaystyle{in{k_{0}\over T}}}
}
\label{twentythree}
\end{equation}

As we have said, it is particularly interesting to evaluate when, on the
phase diagram,
the mass of the sigma becomes
smaller than twice the pion mass, so that it cannot kinematically decay
into its main decay channel.
The phase diagram is now of a more complicated structure than in absence of
chemical potential.
As we know, we expect a tricritical point $(\mu_{t},T_{t})$
separating the region
of second order transition from that
of first order. We thus have to consider different cases.
For $\mu>\mu_{t}$, by letting $T$ grow , one passes through the first order
branch of the critical line.
For $\mu<\mu_{t}$, increasing $T$ one passes through the second order branch.

We recall that of the two parameters present in the coupling,
$\zeta$ is left undetermined and can be only constrained within
some limited region from qualitative considerations.
These are constraints on the critical values of the chemical
potential at very low temperatures, which suggest
to locate this parameter in the range
$\zeta\simeq 0.1-0.3$ to get reasonable values for the critical quark
densities\cite{masstmu}.
At the same time the location of the tricritical point
as well as the appearance of a minimum in the origin (namely
the line of second order phase transitions and its continuation
at low temperatures) are very few affected by the choice
of this parameter. To be specific, the coordinates of the
tricritical point $(\mu_{t},T_{t})$ (in $MeV$) change from
$(76,92)$ for $\zeta=0.1$ to $(75,91)$ for $\zeta=0.3$.

To evaluate $M^2_{\sigma}/M^2_{\pi}$ in Eq.\ (\ref{twentythree})
we can perform the integration
in $k_{0}$ by using the Cauchy theorem and summing up the series;
then, through a numerical calculation we can finally perform
the integration left.
Typical behaviours of the ratio $M_{\sigma}/M_{\pi}$
at fixed $\mu$ and growing $T$ for $\mu<\mu_{t}$
and $\mu>\mu_{t}$ are shown in Fig. 1 and Fig. 2.
We have chosen $\zeta=0.1$, but almost the same
pictures show for other values of $\zeta$ in the range
$0.1-0.3$.
Thus Fig. 1 is for $\mu=70 MeV (<\mu_{t})$, whereas
Fig. 2 is for $\mu=120 MeV (>\mu_{t})$.
The current quark mass is set to
$m=(m_{u}+m_{d})/2=9.5 MeV$.
The little jump visible
in Fig. 2, which occurs after the $\sigma$ mass has
decreased below twice the pion mass, is due to the jump
in the fermion condensate when crossing the line of first
order phase transitions. The size of this jump increases
for higher values of the chemical potential, namely by going far
from the tricritical point.

\section{Conclusions}

Clear-cut experimental signatures for the sigma meson in low energy QCD
phenomenology are made
difficult by its expected large width and by complicated unitarity
contributions and
related distortions. We have examined here the picture emerging at finite
temperatures and densities.
In the simplest model with massless $u$ and $d$ quarks the pion and the
sigma are degenerate
in mass only as long as chiral symmetry holds. At zero temperature and in
absence of
chemical potential, chiral symmetry is broken, the pion is a goldstone of
vanishing
mass, and  the sigma has has a finite mass from the chiral condensate. In
the temperature-density
phase diagram one expects that, when approaching, from the chirally broken
phase, the line of chiral
restoration, the pion mass increases and the sigma mass decreases, and
finally the two particles
become mass degenerate after passing into the region of chiral restoration.
Before
reaching the transition line, within a certain strip touching such a line,
the sigma will not have
phase space left to decay into two pions. The dominant decay channel will
be that into two photons,
and in such ideal conditions the signal from a sigma will be a narrow bump
in the two-photon
invariant mass. The ideal picture will be quantitatively changed  because
of the finite $u$ and $d$
quark masses and the role of the strange quark, but the qualitative feature
of a narrow sigma
resonance is expected to remain there even after inclusion of these
effects. It is too early
to say whether a possible experimental signature for the chiral restoring
transition may be inferred from such behaviour. On the other hand
determination of
the narrow two-photon bump would be of great theoretical interest within
the chiral picture.

The present calculation relies on the composite operator formalism applied
to QCD at finite
temperature and chemical potential. It uses the same QCD model parameters
that had been derived
in previous works, using the same formalism, for the masses and decay
constants of the pseudoscalar
mesons. The  experimental inputs were the mass and decay constant of the
pion, the charged kaon mass
and the electromagnetic mass difference between the neutral and the charged
kaon.
The effective potential is consistently renormalized to derive expressions for
the masses of the scalar and pseudoscalar mesons at finite temperature and
chemical potential. We
have limited the calculation to the phase region where quark-antiquark
condensates
dominate (no quark-quark condensates). The ratio of the sigma mass to
the pion mass is then
calculated within the composite operator formalism for finite temperatures and
densities. We recall that the same model employed here had produced
in the plane of temperature and chemical potential a second order
transition for low chemical potentials and one of first order beyond some
critical
value of the chemical potential (corresponding to a so-called tricritical
point). This feature
induces a related discontinuity in the ratio
 of pion to sigma mass when going beyond the first order transition, that
is for higher densities.
As to the determination of the region in the phase diagram where the sigma
is  no longer
kinematically able to decay
into two pions, we have calculated it quantitatively within the model and
found that, as expected,it
is indeed confined to a small strip near the transition line, and this
independently of
of the type of transition. We have discussed in detail the dependence on
the thermal parameters and
drawn the expected phase diagram behaviours.

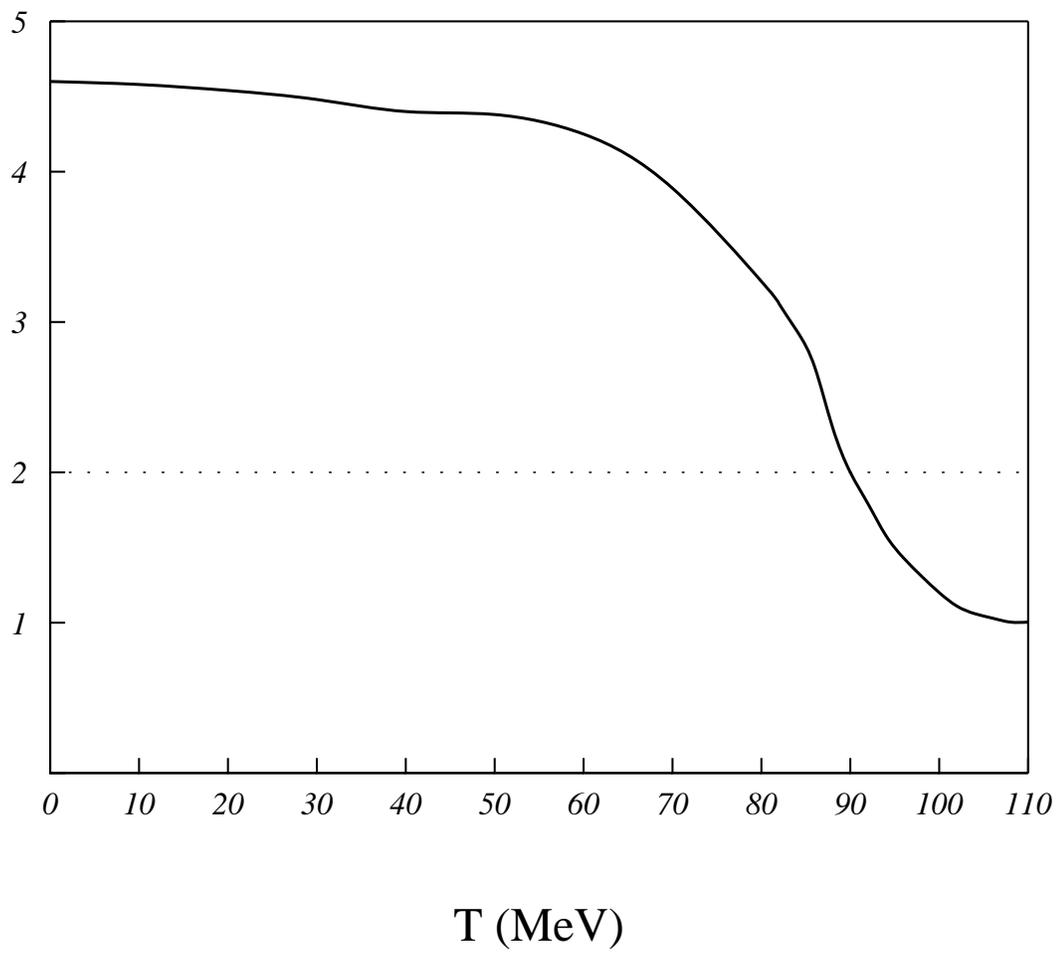
\begin{figure}
\caption{Plot of $M_{\sigma}/M_{\pi}$ versus $T$
for $m=9.5~MeV$.
The chemical potential is fixed at $\mu=70 MeV$,
which is below the value $\mu=\mu_{t}=76MeV$ at the
tricritical point, for $\zeta=0.1$.
\label{fig1}}
\end{figure}

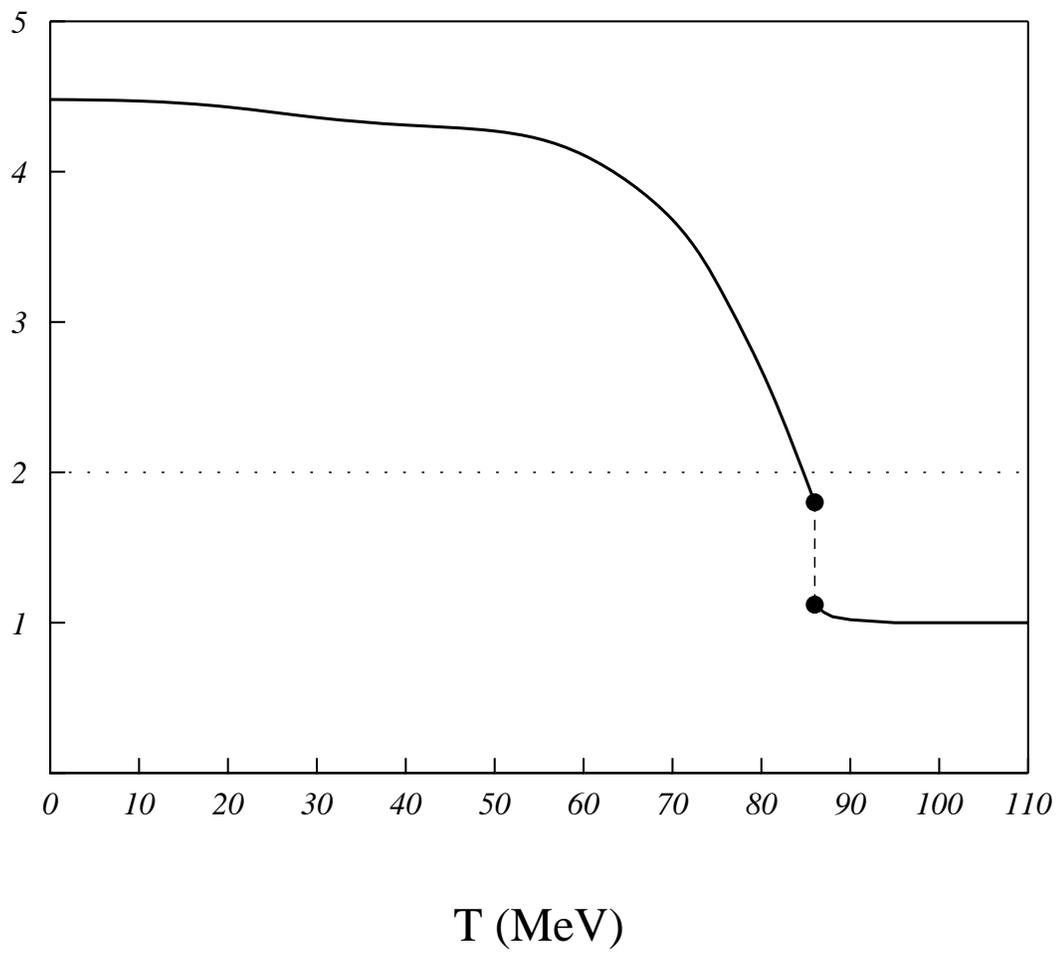
\begin{figure}
\caption{Plot of $M_{\sigma}/M_{\pi}$ versus $T$
for $m=9.5~MeV$.
The chemical potential is fixed at $\mu=120 MeV$,
which is above the value $\mu=\mu_{t}=76MeV$ at the
tricritical point, for $\zeta=0.1$.
\label{fig2}}
\end{figure}

\end{document}